\documentclass[%
 preprint, %linenumbers,
 amsmath,amssymb,
 aps, physrev,
]{revtex4-2}

\usepackage{graphicx}% Include figure files
\usepackage{dcolumn}% Align table columns on decimal point
\usepackage{bm}% bold math
\usepackage{float}
\usepackage[utf8]{inputenc}
\usepackage[T2A, T1]{fontenc}
\usepackage[russian,english]{babel}
\usepackage{mathdots}
\usepackage{makecell}

\begin{document}

\preprint{arXiv preprint}

\title{\textbf{Symmetry-Protected Lossless Modes in Dispersive Time-Varying Media} 
}% 

\author{Calvin M.~Hooper}
 \email{Contact author: ch1122@exeter.ac.uk}
\author{James R.~Capers}
\author{Ian R.~Hooper}
\author{Simon A.R.~Horsley}
\affiliation{%
	Department of Physics and Astronomy\\
	University of Exeter,
    Stocker Road,\\
	Exeter, UK
}%

\date{\today}% It is always \today, today,
             %  but any date may be explicitly specified

\begin{abstract}
	We give an exact application of a recently developed, operator-based theory of wave propagation in dispersive, time--varying media.  Using this theory we find that the usual symmetry of complex conjugation plus changing the sign of the frequency, required for real valued fields, implies that the allowed propagation constants in the medium are either real valued or come in conjugate pairs.  The real valued wave numbers are only present in time--varying media, implying that time variation leads to modes that are free from dissipation, even in a lossy medium.  Moreover, these symmetry-unbroken waves lack a defined propagation direction. This can lead to a divergent transmission coefficient when waves are incident onto a finite, time--varying slab.  The techniques used in this work present a route towards further analytic applications of this operator formalism.
% \begin{description}
% \item[Usage]
% Secondary publications and information retrieval purposes.
% \item[Structure]
% You may use the \texttt{description} environment to structure your abstract;
% use the optional argument of the \verb+\item+ command to give the category of each item. 
% \end{description}
\end{abstract}

%\keywords{Suggested keywords}%Use showkeys class option if keyword
                              %display desired
\maketitle

%\tableofcontents

\section{\label{introduction}Introduction}

The behaviour of light in time-varying materials has lately attracted significant interest~\cite{galiffi2022photonics}, with the promise of novel phenomena including: time-reversal of temporally reflected waves~\cite{morgenthaler1958velocity, mendoncca2003temporal, vezzoli2018optical}; a temporal Brewster angle~\cite{pacheco2021temporal}; analogue Hawking radiation~\cite{pendry2021gain, horsley2023quantum, horsley2024travelling}; non-Hermitian spectral manipulation~\cite{koutserimpas2018nonreciprocal}; non-reciprocal devices without magnetism~\cite{sounas2017non}; or exceptional points in spatially homogeneous media~\cite{Koutserimpas2020}. Unfortunately, analytic methods in such media remain limited. This is in contrast to static media, where, despite a still very limited number of exact solutions, the salient features are largely understood. The addition of temporal modulation significantly complicates matters. In static media, passing through a spatially homogeneous medium provides only a phase shift.  Meanwhile, passing through a spatially homogeneous but time-varying material in general leads to frequency conversion.  At a time--varying interface, this conversion then implies a coupling between multiple angles of reflection, not just forwards- and backwards-moving waves. Finally, even in spatially homogeneous materials, simultaneous temporal variation and dispersion provides two competing ways for a pulse to disperse, significantly complicating any analytic solution.

\begin{figure}
    \centering
    \includegraphics[width=1\textwidth]{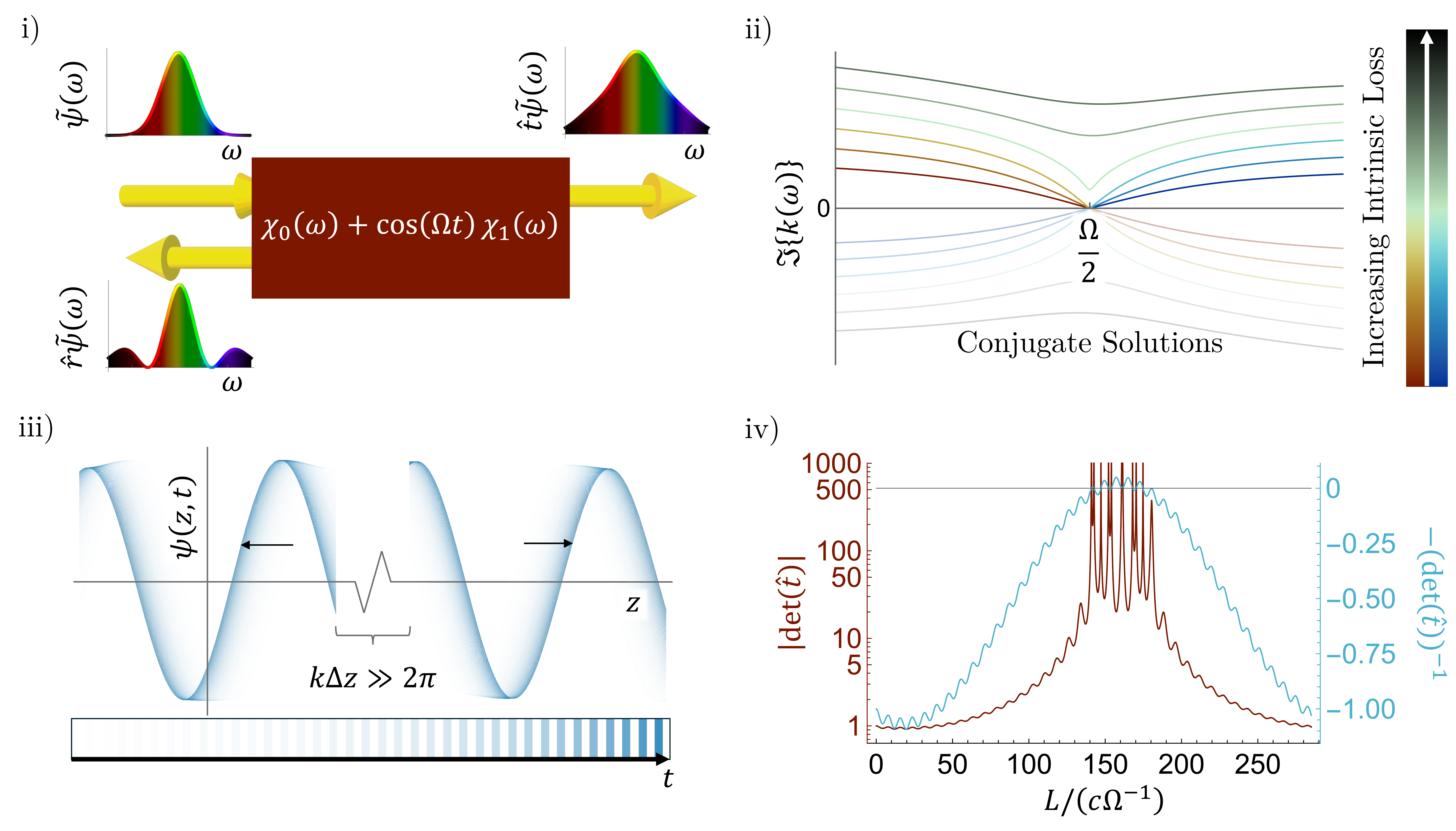}
    \caption{\textbf{Features of propagation through a dispersive, time--varying slab:} i) In this paper we calculate the propagation characteristics of waves in a 1D slab with a dispersive susceptibility, varying in time with an angular frequency \(\Omega\). Reflection and transmission spectra from this system can be understood in terms of \(\widehat{r}\) and \(\widehat{t}\) operators (respectively) acting on the frequency spectrum \(\widetilde{\psi}\left(\omega\right)\) of an incoming wave. ii) Protection from loss. The loss experienced by waves in a time-varying material, with an incident frequency of \(|\omega| = \frac{\Omega}{2}\). For a range of values of the material loss, incident waves with \(\omega = \frac{\Omega}{2}\) experience no absorption. iii) Adirectionality.  Here \(\psi(z,t)\) is an example wave in this medium (incident with \(|\omega| = \frac{\Omega}{2}\)), with its history indicated as a fading of each plotted line (see legend inset). The wave appears to entirely switch direction between the regions shown, a fact confirmed by the local power flux, shown in black arrows. iv) Determinant of the transmission operator (brown), seen to reach arbitrarily large values for particular cavity lengths. The changes in sign of the reciprocal determinant (blue) show that this is a true divergence. Parameters for this model are provided in section 2.}
    \label{fig: SummaryFigure}
\end{figure}

One successful approximation to space--time varying media is given in~\cite{horsley2023quantum, horsley2024travelling}, neglecting reflection and dispersion through assuming a constant impedance, with a real, frequency independent refractive index, (\(\mu(z,t) \propto \varepsilon(z,t)\)).  In practice, however \(\varepsilon\) is typically easier to control than \(\mu\), especially in the regime of recent optical experiments~\cite{bohn2021all, bohn2021spatiotemporal, tirole2022saturable, tirole2023double, zhou2020broadband, vezzoli2018optical}.  For this reason,~\cite{horsley2023eigenpulses} recently suggested an operator formalism for investigating media with time-variation, dispersion, and even sharp spatial boundaries. In this approach, the square wavenumber \(K^{2} = \frac{\omega^{2}}{c^{2}}(1 + \chi)\) is promoted to an operator acting on the frequency spectrum of the wave, with an accompanying operator dispersion relation. In this language, most time-varying results, such as slab transmission and reflection coefficients (Fig. \ref{fig: SummaryFigure}.i), can be derived in the same way as their static counterparts, but are promoted to operators that act on the frequency spectrum of the wave.

The downside with this formalism is that these operators are typically infinite in extent, precluding exact analyses. In existing work, numerical results have been obtained through truncating these operators to a finite subspace, the properties converging for large enough subspaces.  Yet exact results remain elusive. In this work we find a class of dispersive, time-varying wave equations with exact solutions.

The supported waves fall into two classes, as also noted in~\cite{MartnezRomero2017}. The first behaves precisely according to standard intuition for static lossy materials, propagating power whilst dissipating energy, leading to finite slab-transmission coefficients. Meanwhile, the second class exhibits neither gain nor loss, lacks a defined propagation direction, and possesses divergent transmission for particular slab lengths (Fig. \ref{fig: SummaryFigure}.ii-iv).

These two classes are rigorously distinguished depending on whether they break a key symmetry of the system. Unsurprisingly, in a real wave equation, fields which are real at a spatial boundary remain real everywhere. This implies a Fourier-space symmetry, henceforth referred to as \(\mathcal{RC}\)-symmetry; formed by the composition of the operators for frequency-reflection (\(\mathcal{R}\widetilde{\psi}(\omega) = \widetilde{\psi}( - \omega)\)) and complex conjugation (\(\mathcal{C}\widetilde{\psi}(\omega) = {\widetilde{\psi}}^{*}(\omega)\mathcal{C}\)), fully characterised in appendix \ref{rc-symmetry}. Spontaneous symmetry breaking characterises the first class, whilst the second possess unbroken \(\mathcal{RC}\)-symmetry.

Since (un)broken symmetry underlies the physics of these classes, their existence is robust to perturbation. As a result, the results and intuition from these exactly solvable models extend to a much wider range of physical systems.

As an example, we consider a Drude metal with a time-varying electron density, demonstrating that, with sufficient modulation, \(\mathcal{RC}\)-unbroken optical waves are supported. This is motivated both by recent experiments in ITO~\cite{bohn2021all, bohn2021spatiotemporal, tirole2022saturable, tirole2023double}, but also simply as a minimal realistic model including loss.

An introduction to our key methods is included in the first two sections, outlining the operator formalism of~\cite{horsley2023eigenpulses}. We then give an example of a solvable model. With these methods in place, we explain the initially counterintuitive phenomena presented above, building from the behaviour of an infinite medium to that of a finite slab.

\section{Waves in the Operator Formalism}\label{waves-in-the-operator-formalism}

We consider the system sketched in Fig. \ref{fig: SummaryFigure}.i: a dielectric slab with a time-varying, dispersive electric susceptibility, which for an incident wave of frequency $\omega$ is assumed to have the form $\chi_{\rm E}(\omega,t)=\chi_{0}(\omega)+\cos(\Omega t)\chi_{1}(\omega)$. We assume normal incidence, propagation along the $z$ axis, and a fixed polarization $\boldsymbol{\mathrm{E}}=\psi\boldsymbol{\mathrm{e}}_{x}$, so that Maxwell's equations can be reduced to $\boldsymbol{\nabla}\times\boldsymbol{\mathrm{H}}=\partial_{z}(\boldsymbol{\mathrm{e}}_{z}\times\boldsymbol{\mathrm{H}})=\boldsymbol{\mathrm{e}}_{x}\partial_t(\epsilon_0\hat{\epsilon}\psi)$ and $\boldsymbol{\nabla}\times\boldsymbol{\mathrm{E}}=\boldsymbol{\mathrm{e}}_{y}(\partial_{z}\psi)=-\partial_t(\mu_0\boldsymbol{\mathrm{H}})$.  Combining these two equations, the wave equation within the modulated slab is given by

\begin{align}
    \frac{\partial^{2}\psi}{\partial z^{2}} = \frac{1}{c^{2}}\frac{\partial^{2}}{\partial t^{2}}\left[\left( 1 + \hat{\chi}_{0}\left( \mathrm{i}\partial_{t} \right) + \cos(\Omega t)\hat{\chi}_{1}\left( \mathrm{i}\partial_{t} \right) \right)\psi\right].
    \label{eq: TimeDomainWaveEquation}
\end{align}

This is a \(1\)D wave equation for the field \(\psi\).  Neglecting the time modulation, this is the usual scalar wave equation for a medium with a frequency-dependent susceptibility \(\chi_{0}(\omega)\).  The time modulation of this susceptibility is proportional to \(\chi_{1}(\omega)\), oscillating with a driving frequency \(\Omega\).  Here we use the form \(\chi_{0}(\omega) = \eta^{- 1}\chi_{1}(\omega) = - \omega_{\mathrm{pl},0}^{2} \left(\omega^{2} + \mathrm{i}\gamma\omega\right)^{-1}\), where $\eta$ is a constant quantifying the contrast of the time modulation.  This choice of permittivity corresponds to the Drude model of a metal with scattering rate \(\gamma\), and a time-varying plasma frequency of \(\omega_{\mathrm{pl}}^{2}(t) = \omega_{\mathrm{pl},0}^{2}\left( 1 + \eta\cos(\Omega t) \right)\), due to e.g. a time-varying electron density.  In the following numerical calculations we work in units where \(\Omega = 1\), \(c = 1\), \(\omega_{\mathrm{pl},0} = 0.3\), \(\gamma = 10^{- 2}\), and \(\eta = 0.2\).

Following the operator formalism~\cite{horsley2023eigenpulses}, Eq. (\ref{eq: TimeDomainWaveEquation}) can be Fourier Transformed by the replacements: $\psi\to\tilde{\psi}$, ${\rm i}\partial_t\to\omega$, and $t\to-{\rm i}\partial_{\omega}$, and can then be written as a square wavenumber operator \(\hat{K}^2\) acting on the frequency spectrum of the wave,
% \begin{equation}
%     \left( \begin{array}{r}
%     \psi \\
%     \mathrm{i}\partial_{t} \\
%     t
%     \end{array} \right) \mapsto \left( \begin{array}{r}
%     \widetilde{\psi} \\
%     \omega \\
%      - \mathrm{i}\partial_{\omega}
%     \end{array} \right).
% \end{equation}
\begin{equation}
    \frac{\partial^{2}\widetilde{\psi}}{\partial z^{2}} = - \frac{\omega^{2}}{c^{2}}\left( 1 + \chi_{0}(\omega) + \cos\left( - \mathrm{i}\Omega\partial_{\omega} \right)\chi_{1}(\omega) \right)\tilde{\psi} = - \hat{K}^{2}\widetilde{\psi}.
    \label{eq: FrequencyDomainWaveEquation}
\end{equation}

In precise analogy with the scalar case, solutions to Eq. (\ref{eq: FrequencyDomainWaveEquation}) may be written in the form

\begin{equation}
    \widetilde{\psi} = e^{\mathrm{i}\hat{K}z}{\widetilde{\psi}}_{+} + e^{- \mathrm{i}\hat{K}z}{\widetilde{\psi}}_{-},
    \label{eq: OperatorFormWaveSolution}
\end{equation}

where the effect of time-variations is now encoded within the exponentials of the operator \(\hat{K}\), acting to mix frequencies in the spectrum of the complex electric field amplitude, \(\widetilde{\psi}(z)\). Although the square root of this operator is ambiguous (i.e. for an $N\times N$ diagonalizable matrix there are $2N$ different roots), whichever root of \(\hat{K}^{2}\) we choose is irrelevant, given that both $+\hat{K}$ and $-\hat{K}$ are included in Eq. (\ref{eq: OperatorFormWaveSolution}), and for simplicity here we take the principal root (see appendix \ref{square-root-choice}).  We note however, that the usual physical interpretation of (\ref{eq: OperatorFormWaveSolution}) as left and right going waves is dependent on this choice of root.  For example, for $\exp(+{\rm i}\hat{K}x)\tilde{\psi}$ to correspond to a purely right going waveform, all eigenvalues of $\hat{K}$ with a positive real part, should be associated with eigenvectors which have non--zero entries in only the positive frequency part of the spectrum.  In many cases it is not possible to find such a root.

\begin{figure}
    \centering
    \includegraphics[width=0.5\textwidth]{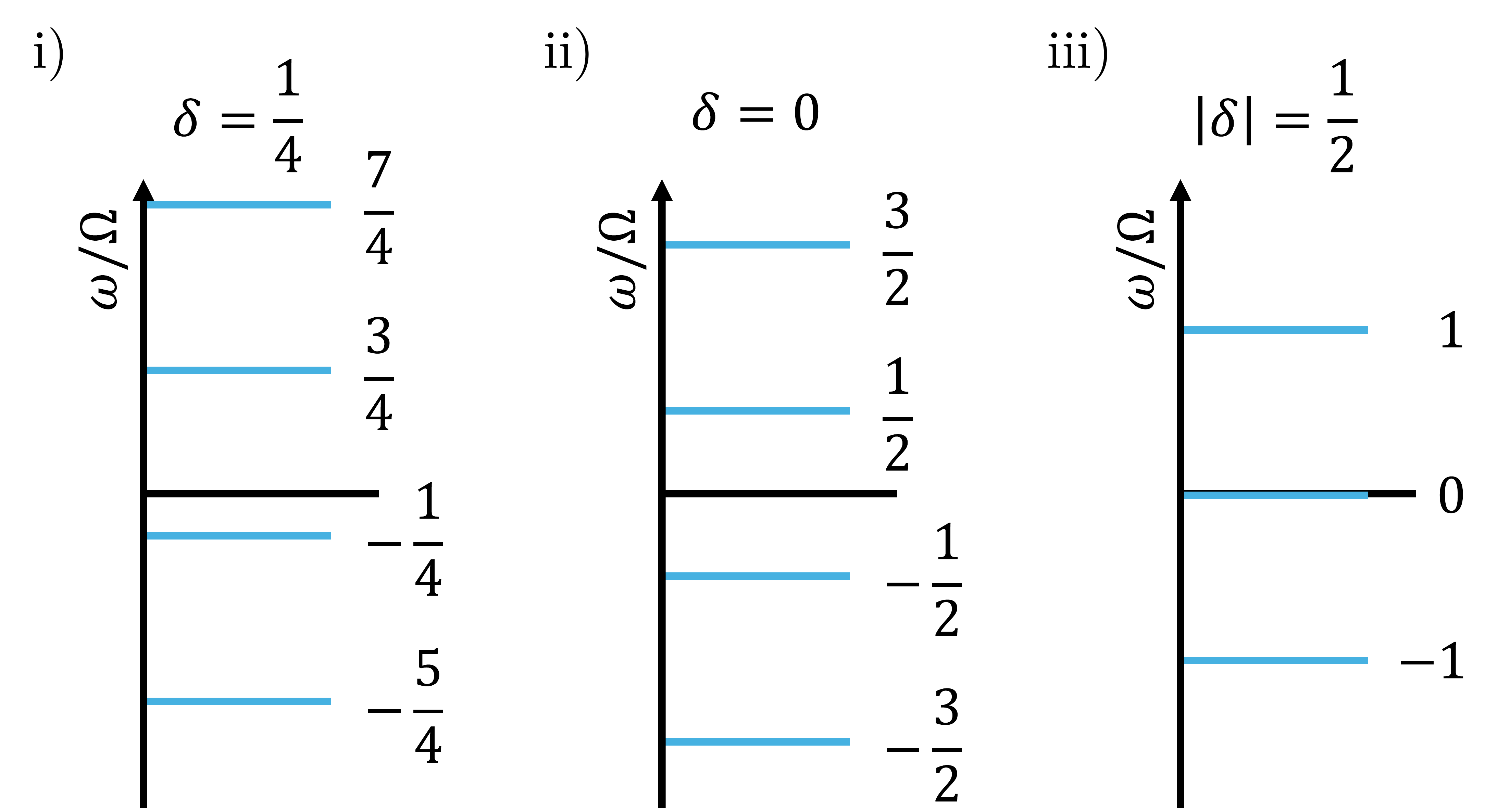}
    \caption{\textbf{Frequency Ladders in sinusoidally modulated media:} Any given central frequency, \(\omega_c=\left( \frac{1}{2} + \delta \right)\Omega\) can only be coupled to frequencies separated by integer multiples of \(\Omega\).  Due to the periodic nature of each ladder, this offset can be taken as \(|\delta| \leq 0.5\). a) For most values of \(\delta\), the corresponding frequency ladder is not (\(\mathcal{RC}\)-)symmetric about \(\omega = 0\). b) For \(\delta = 0\), the most important case for this paper, the ladder is (\(\mathcal{RC}\)-)symmetric about \(\omega = 0\). c) For \(\delta = \pm 0.5\), the ladder contains only integer multiples of \(\Omega\). Since the ladder is identical for \emph{both} \(\delta = \pm 0.5\), we denote it with \(|\delta| = 0.5\).}
    \label{fig: FrequencyLadders}
\end{figure}

One key feature immediately arises as a result of having chosen a sinusoidal time-dependence: \(K^{2}\) only couples frequencies separated by integer multiples of \(\Omega\). This follows from noticing that the cosine (or sine) operator, \(\cos\left( - \mathrm{i}\Omega\partial_{\omega} \right)=\frac{1}{2}\left[\exp(\Omega\partial_{\omega})+\exp(-\Omega\partial_{\omega})\right]\) is just a sum of displacement operators that serve to move the spectrum of the field up or down in frequency $\omega\to\omega\pm\Omega$.  Therefore we can choose some central frequency \(\omega_c=\left( \frac{1}{2} + \delta \right)\Omega\), and restrict our attention only to frequencies some multiple of $\Omega$ away from $\omega_c$.  This ladder of frequencies is sketched in Fig. \ref{fig: FrequencyLadders}.  The full operator, \(\hat{K}^{2}\) can thus be split into a tensor sum of operators \(\hat{K}_{\delta}^{2}\) (see appendix \ref{operator-definitions}), each responsible for coupling within a discrete ladder of frequencies centred around $\omega_c$.

\section{Truncating the wavenumber operator}\label{truncation-of-k2}

\begin{figure}
    \centering
    \includegraphics[width=0.9\textwidth]{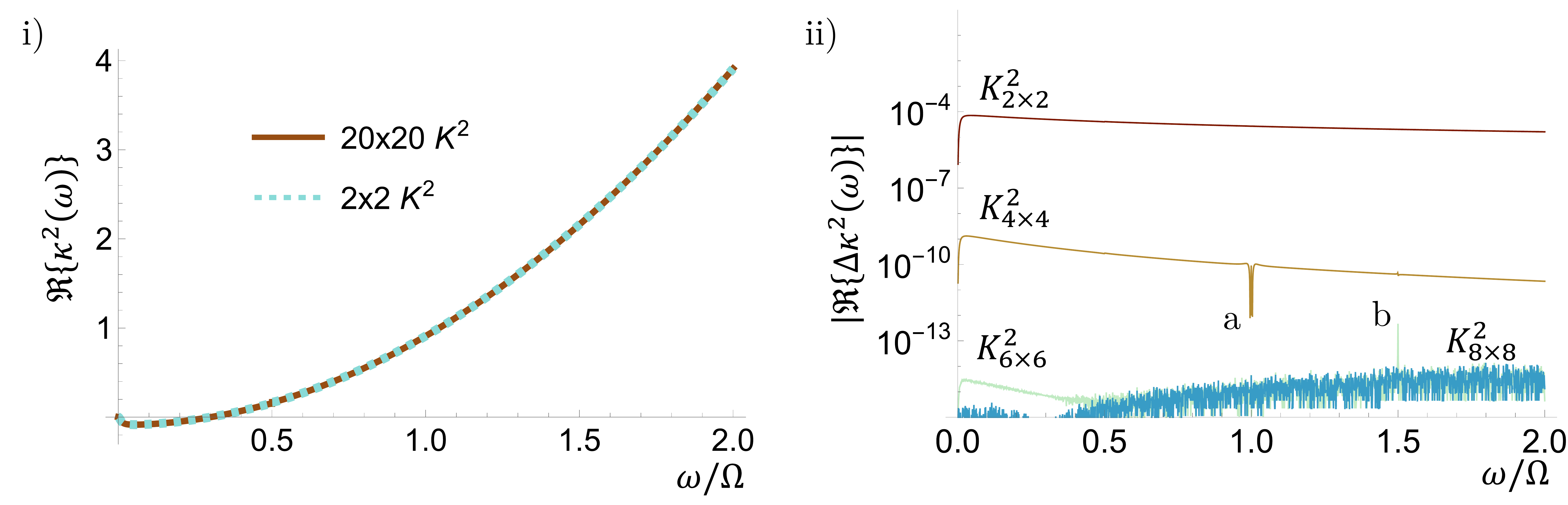}
    \caption{\textbf{Approximating} \(\mathbf{K}_{\mathbf{\delta}}^{\mathbf{2}}\) \textbf{with a truncated matrix:} i) The dispersion relation of \(\hat{K}^{2}\) when truncated to \(2\)-by-\(2\) and \(20\)-by-\(20\) matrices. The dispersion relation is generated by calculating the eigenvalues of the $\hat{K}^{2}$ matrix for different incident frequencies, $\omega$ (and thus different ladders, as shown in Fig. \ref{fig: FrequencyLadders}).  The value of \(\kappa^{2}\left(\omega\right)\) is obtained by normalising the eigenvalue of \(K^{2}\) with respect to \(\frac{\Omega}{c}\).  We compare the same eigenvalue in the $20\times20$ case to that in the $2\times2$ case (where the real part is the same for both eigenvalues).  ii) Convergence of increasingly large truncations to that of the \(20\)-by-\(20\) approximation.  By the time we consider an \(8\)-by-\(8\) matrix, finite computational precision is a more significant error than that due to the truncation. Highlighted features a and b are discussed in the main text.}
    \label{fig: TruncationError}
\end{figure}

We now consider e.g. a wave incident onto our medium with a fixed value of $\delta$ (frequency $\omega_c$), thus picking out one of the aforementioned sub--operators, \(K_{\delta}^{2}\).  Although this means we can reduce the operator down to a matrix that acts only on one of the frequency ladders shown in Fig. \ref{fig: FrequencyLadders}, this matrix is still infinite-dimensional, making exact solutions difficult to find.  However, by tuning the coupling term $\chi_{1}$ to zero at certain frequencies, the spectrum of the field can be confined to a finite number of frequencies. In these situations, \(K_{\delta}^{2}\) is a finite-dimensional matrix, and exact solutions to the wave equation (\ref{eq: FrequencyDomainWaveEquation}) can be obtained.  These truncated matrices can also be viewed as an approximation to the operators in any system (as is required by numerical computations, such as those of \cite{horsley2023eigenpulses}). In Fig. \ref{fig: TruncationError}, convergence of the dispersion relation is shown as the dimensionality of \(K^2\) is increased.

As an aside, note that the rate of convergence is sensitive to coupling within the truncated matrix. For instance, in Fig. \ref{fig: TruncationError}.ii.a, second order coupling through the high response around (but not exactly at) \(\omega=0\) produces rapid variations in \(\kappa\left(\omega\right)\), resulting in an improved approximation. By contrast, the reduced accuracy around Fig. \ref{fig: TruncationError}.ii.b is physical in origin; the proximity of a momentum bandgap enhances the effect of coupling to higher frequencies on the resulting eigenvalues. Fortunately, a sufficiently high dimensional truncation resolves these problems (as is seen in the bounded error for \(K_{8\times 8}^{2}\)).

A more interesting fact is that matrices as small as \(2\)-by-\(2\) still provide accurate results.  Indeed, note that for some choices of material dispersion, such a truncated matrix can also produce \emph{exact} results.  For example, in Eq. (\ref{eq: FrequencyDomainWaveEquation}) we can see that the off--diagonal terms in $\hat{K}^2_{\delta=0}$ are given by $\frac{\omega^2}{c^2}\cos{\left(-\mathrm{i}\Omega \partial_{\omega}\right)}\chi_{1}\left(\omega\right)$, with $\omega=(n+\frac{1}{2})\Omega$.  If the dispersion of the medium is such that $\chi_{1}(3\Omega/2)=\chi_{1}^{\star}(-3\Omega/2)=0$ then there is no coupling between $\omega=\pm\Omega/2$ and the neighbouring elements on the frequency ladder, $\omega=\pm 3\Omega/2$, allowing us to consider a two dimensional subspace containing only the frequencies $\pm\Omega/2$.

We now assume $\delta=0$ and make the above truncation to a $2\times2$ matrix, focusing our attention on the behaviour of the eigenvectors (henceforth referred to as eigenpulses, following \cite{horsley2023eigenpulses}) of \(K^{2}\) with---what we call here---\(\mathcal{RC}\) symmetry, i.e. the modes are invariant under simultaneous reflection $\omega\to-\omega$ and complex conjugation, as discussed in the introduction (see appendix \ref{rc-symmetry}).  The truncated matrix is given by

\begin{equation}
    \hat{K}_{\delta=0}^{2} = \left( \frac{\Omega}{2c} \right)^{2}
    \begin{pmatrix}
        1 + \chi_{0}^{*} & \frac{1}{2}\chi_{1} \\
        \frac{1}{2}\chi_{1}^{*} & 1 + \chi_{0}
    \end{pmatrix},
    \label{eq: TruncatedCentralKSquared}
\end{equation}

where \(\chi_{0,1}=\chi_{0,1}(\Omega/2)\).  So far our assumptions amount to considering two frequencies symmetrically spaced around $\omega=0$, the time modulation converting positive frequency to negative frequency and vice versa, as in the case of both time reflection~\cite{galiffi2022photonics} and parametric amplification, where the driving frequency must be twice the oscillation frequency.  Eq. (\ref{eq: TruncatedCentralKSquared}) is also analogous to the theory of diffraction from a spatially periodic index profile~\cite{petit2013}, for a wave-vector close to the Brillouin zone boundary, when we can also restrict our attention to two modes.

An important property of the operator \(\hat{K}_{\delta=0}^{2}\) is the aforementioned \(\mathcal{RC}\)-symmetry, i.e. invariance under a combined \(180{^\circ}\) rotation of the matrix elements, and complex conjugation, i.e.

\begin{equation}
    (\sigma_x\hat{K}_{\delta=0}^2\sigma_x)^{\star}=\hat{K}_{\delta=0}^2.\label{eq:symmetry}
\end{equation}

Similar to the theory of parity--time symmetry~\cite{bender2007making,bender2019}, this symmetry implies that the eigenvalues $k_{\pm}$ of $\hat{K}_{\delta=0}^2$ are either real, or come in complex conjugate pairs.  The symmetry (\ref{eq:symmetry}) implies that if $\tilde{\psi}_{+}$ is an eigenvector with eigenvalue $k_+$, then $(\sigma_{+}\tilde{\psi}_{+})^{\star}$ is also an eigenvector, but with eigenvalue $k_+^{\star}$.  Therefore our two eigenpulses are either invariant under \(\mathcal{RC}\)-symmetry, and have a completely real propagation constant (we dub these `lossless eigenpulses'), or break \(\mathcal{RC}\)-symmetry and are converted into one another by this symmetry operation.

There are two important things to note here.  Firstly, without time variation, the operator (\ref{eq: TruncatedCentralKSquared}) is diagonal with complex conjugate pairs of diagonal entries, and thus static media \emph{always} corresponds to the \(\mathcal{RC}\)-symmetry--broken case.  The aforementioned lossless eigenpulses are thus purely a feature of time modulated systems.  Although these zero--loss modes are analogous to the modes at the band edge in a spatially periodic medium, where the wave switches from propagation to decay, there is a crucial difference here.  In a lossy spatially periodic medium, the propagation constant is \emph{always} complex valued for frequencies both inside and outside the bandgap.  However here we have a lossy system where the dissipation can be completely suppressed by the time modulation, a fact that is guaranteed by an extremely general symmetry. 

Secondly, as we just mentioned, this is a general symmetry: the above argument doesn't at all rely on the dimension of our $2\times2$ truncated matrix, but holds for any size matrix, with the Pauli matrix replaced with the anti--diagonal exchange matrix.  Again, the eigenvectors with unbroken \(\mathcal{RC}\)-symmetry have a completely real propagation constant.

% It can be shown without much difficulty that, aside from a single mode (corresponding, in terms of electromagnetism, to a constant \(\mathbf{D}\) field), any \(\mathcal{RC}\)-unbroken eigenpulses must belong to the \(\delta=0\) frequency ladder. As a result, in the remainder of the paper, we find it sufficient to consider the properties of \({\overline{K}}_{0}^{2}\).
% To illustrate this truncation, consider \(K_{0}^{2}\), as defined in \ref{eq: FrequencyDomainWaveEquation}, and in particular its off-diagonal terms, given by

% \begin{equation}
%     K_{\mathrm{off-diagonal}}^{2}=\frac{\omega^2}{c^2}\cos{\left(-\mathrm{i}\Omega \partial_{\omega}\right)}\chi_{1}\left(\omega\right).
%     \label{eq: OffDiagonalK2}
% \end{equation}

% By setting \(\chi_{1}\left(\omega\right)=0\) for some \(\omega\), and that frequency become irrelevant for determining the behaviour of other frequencies. Hence, by setting \(\chi_{1}\left(\frac{3 \Omega}{2}\right)=\chi_{1}^{*}\left(-\frac{3 \Omega}{2}\right)=0\), the coupling between \(\pm\frac{\Omega}{2}\) becomes independent of any other frequencies, and is described by the truncated matrix

\section{Waves in Infinite Time-Varying Media}\label{dispersion-in-time-varying-media}

\subsection{Lossless Eigenpulses}\label{lossless-band-edges}

% In this section, we prove three key properties associated with \(\mathcal{RC}\)-unbroken eigenpulses. Namely, that their existence implies a bandgap of unstable wavenumbers; that waves at the edges of these bandgaps have identically \(0\) loss; and that these waves cannot be described as either leftwards- or rightwards-moving.
With \({\hat{K}}_{\delta=0}^{2}\) defined in (\ref{eq: TruncatedCentralKSquared}), examples of \(\mathcal{RC}\)-unbroken eigenpulses are not difficult to find. The eigenvalues of this $2\times 2$ operator equal

\begin{equation}
    k_{\pm}^{2} = \left( \frac{\Omega}{2c} \right)^{2}\left( 1 + \mathfrak{R}\left\{ \chi_{0} \right\} \pm \sqrt{\frac{1}{4}\left| \chi_{1} \right|^{2} - \left( \mathfrak{I}\left\{ \chi_{0} \right\} \right)^{2}} \right).
    \label{eq: K02Eigenvectors}
\end{equation}

\begin{figure}
    \centering
    \includegraphics[width=1\textwidth]{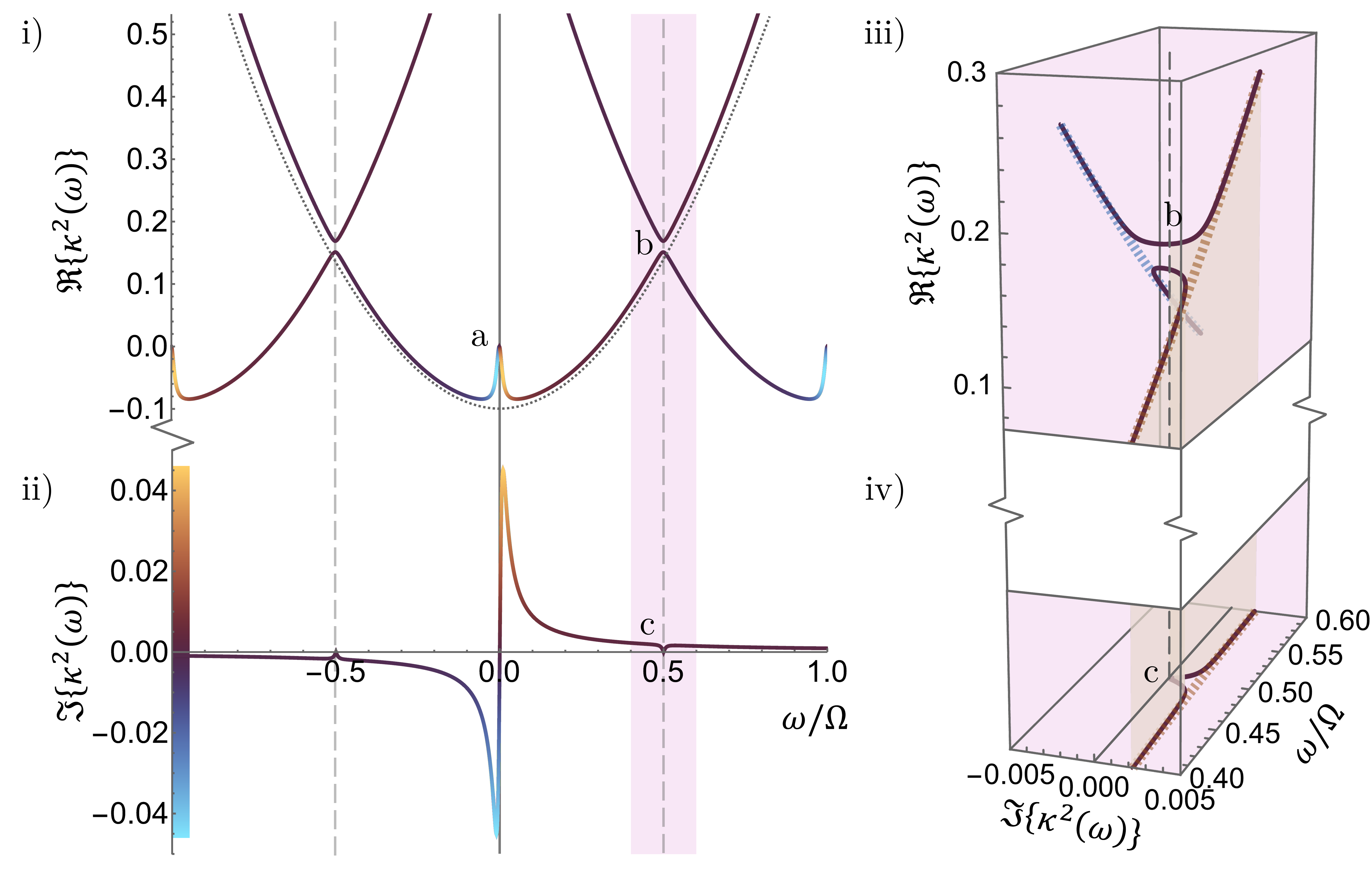}
    \caption{\textbf{Loss Protection at} \(\boldsymbol{k}^{\mathbf{2}}\)\textbf{-Bandgaps:} Complex dispersion relation of a metal with an oscillating plasma frequency, presented with the dimensionless square wavenumber \(\kappa^{2}(\omega)=\frac{c^{2}}{\Omega^{2}}k^{2}(\omega)\), as defined in Fig. \ref{fig: TruncationError}.\\
    i) Line plot of \(\mathfrak{R}\left\{ \kappa^{2}(\omega)\right\}\), with \(\mathfrak{I}\left\{ \kappa^{2}(\omega)\right\}\) represented in colour (according to the colour bar provided in (ii)), presented as a periodic Brillouin zone (with zone boundaries denoted by grey dashed lines). The dispersion relation in the extended Brillouin zone is found by selecting, at each band-edge, the branch highlighted by the grey dotted line.
    Metallic behaviour dominates near (a), while lossless behaviour emerges at the Brillouin zone boundary (b).
    % At very low frequencies, metallic behaviour dominates, seen through the large imaginary parts around (a). \(\mathcal{RC}\)-symmetry ensures symmetry of the real part, and antisymmetry of the imaginary part (blue \(\leftrightarrow\) orange). For sufficiently strong modulation and low loss, a bandgap opens up at (b). Due to the periodicity of the Brillouin zone, the same symmetry properties hold about this point. Thus, at a bandgap the imaginary part of the dispersion vanishes, corresponding to a lossless eigenpulse at the band edge. 
    ii) \(\mathfrak{I}\left\{ \kappa^{2}(\omega)\right\}\), plotted explicitly over the extended Brillouin zone. As in (i), loss vanishes at the Brillouin-zone boundary (c).\\
    iii) A 3D plot of the complex dispersion relation (viewed in a periodic Brillouin zone), about the band-edge (b) allows the properties of (i) and (ii) to be viewed geometrically. For strong modulation and low loss, the unmodulated dispersion relation (light dotted lines) couples to form a bandgap. Here, the combined \(\mathcal{RC}\)- and Brillouin-symmetry enforces rotational symmetry about the axis (b), and thus that \(\mathfrak{I}\left\{ \kappa^{2}\left(\frac{\Omega}{2}\right) \right\}=0\).
    iv) The imaginary part of the dispersion relation near the bandgap (c), in the extended Brillioun zone, projected from (iii).}
    \label{fig: Bandstructure}
\end{figure}

As proven in appendix \ref{rc-symmetry}, \(k_{\pm}^{2}\mathbb{\in R}\) is a necessary and sufficient condition for an eigenpulse to be \(\mathcal{RC}\)-unbroken.  In terms of material parameters, this implies the existence of a lossless eigenpulse whenever \(\left| \chi_{1} \right| > 2\left| \mathfrak{I}\left\{ \chi_{0} \right\} \right|\). This inequality sets up an important dichotomy: coupling between positive and negative frequencies acts to preserve modes with \(\mathcal{RC}\)-symmetry, whilst non-time-varying losses act to break it.  When these two effects are balanced there is an exceptional point with \(\left| \chi_{1} \right| = 2\left| \mathfrak{I}\left\{ \chi_{0} \right\} \right| \neq 0\) (or a simple degeneracy for \(\left| \chi_{1} \right| = 2\left| \mathfrak{I}\left\{ \chi_{0} \right\} \right| = 0\)). The exceptional point in this \(\chi\) operator can be directly related to the exceptional point demonstrated in~\cite{Jin2024} for a single oscillator, but generalised to a medium model. Running this calculation for a range of \(\delta\) allows the dispersion \(k^{2}(\omega)\) to be calculated for this system (Fig. \ref{fig: Bandstructure}), where \(\mathcal{RC}\)-unbroken modes lead to a wavenumber bandgap.  This is shown in Fig. \ref{fig: Bandstructure}.a.

As mentioned above, a feature of \(\mathcal{RC}\)-unbroken eigenpulses is that they possess a real propagation constant \(k_{\pm}^{2}\), even when loss is present in a system (\(\mathfrak{I}\left\{ \chi_{0} \right\} > 0\)). Physically, this can be interpreted with regard to the even balance of positive and negative frequencies in \(\mathcal{RC}\)-unbroken eigenpulses (\(\mathcal{RC}\)-symmetry implies every Fourier component  of the field $\widetilde{e}$ obeys \(\widetilde{e}(\omega) = {\widetilde{e}}^{*}( - \omega)\), and thus \(\left| \widetilde{e}(\omega) \right| = \left| \widetilde{e}( - \omega) \right|\)). As in the Zel'dovich effect\cite{PismaZhETF.14.270, faccio2019superradiant, cromb2020amplification}, a lossy susceptibility at positive frequencies produces gain when applied to negative frequencies. Thus, we find that this even balance of frequencies prevents any loss in \(\mathcal{RC}\)-unbroken modes.

Both the existence of bandgaps, and this protection from loss can also be viewed geometrically. As in spatial crystals, bandgaps can be interpreted based on coupling between two neighbouring Brillouin zones (Fig. \ref{fig: Bandstructure}.a). Since this corresponds to coupling between positive and negative frequencies, \(\mathcal{RC}\)-symmetry of the dispersion relation ensures that the real (imaginary) part of the dispersion relation is (anti)symmetric about \(\delta = 0\), where the coupling occurs. This symmetry ensures that sufficient coupling to produce a bandgap simultaneously implies zero loss at the boundary. Furthermore, whilst in spatial crystals the presence of a bandgap leads to zero group velocity, here we observe a point of infinite group velocity at the bandgap, as discussed in~\cite{MartnezRomero2017}.

\subsection{Adirectional Propagation}\label{adirectional-propagation}

Whilst the eigenvalues associated with \(\hat{K}_{\delta=0}^{2}\) may be intuitive, the form of its eigenvectors remains somewhat obfuscated. To investigate the differences between \(\mathcal{RC}\)-broken and -unbroken eigenpulses, we consider two particularly simple examples of \(\hat{K}_{\delta=0}^{2}\), described in Table~\ref{tab: ExemplarK2Matrices}. These particular examples come with the caveat that they do not provide an example of isolated exceptional points.  This is also the reason their eigenvectors are so simple.

\begin{table}
    \centering
    \begin{tabular}{|c|c|c|}
        \hline
            &
            i) \(\mathcal{RC}\)-broken &
            ii) \(\mathcal{RC}\)-unbroken \\
        \hline
        
        &&\\
            Constraints &
            \(\chi_{1} = 0;\ \mathfrak{I}\left\{ \chi_{0} \right\} \neq 0\) &
            \(\mathfrak{I}\left\{ \chi_{0} \right\} = 0;\ \mathfrak{I}\left\{ \chi_{1} \right\} = 0;\ \chi_{1} \neq 0\) \\
        
        &&\\
            \(\hat{K}_{\delta=0}^{2}\) &
            \(\hat{K}_{\delta=0}^{2} = \left( \frac{\Omega}{2c} \right)^{2}
            \begin{pmatrix}
                1 + \chi_{0}^{*} & 0 \\
                0 & 1 + \chi_{0}
            \end{pmatrix}
            \)
            & \(\hat{K}_{\delta=0}^{2} = \left( \frac{\Omega}{2c} \right)^{2}
            \begin{pmatrix}
                1 + \chi_{0} & \frac{1}{2}\chi_{1} \\
                \frac{1}{2}\chi_{1} & 1 + \chi_{0}
            \end{pmatrix}\) \\
        
        &&\\
            Eigenvectors &
            \makecell{
                \(\mathbf{e}_{-} = \mathcal{RC}{\mathbf{e}}_{+} = \begin{pmatrix}
                    1 \\
                    0
                \end{pmatrix}\)
                \\
                \(\mathbf{e}_{\mathbf{+}} = \mathcal{RC}{\mathbf{e}}_{-} = \begin{pmatrix}
                    0 \\
                    1
                \end{pmatrix}\)
            }
            &
            \makecell{
                \(\mathbf{e}_{+} = \mathcal{RC}\mathbf{e}_{+} = \frac{1}{2}
                \begin{pmatrix}
                    1 \\
                    1
                \end{pmatrix}\)
                \\
                \(\mathbf{e}_{\mathbf{-}} = \mathcal{RC}\mathbf{e}_{-} = \frac{1}{2} \begin{pmatrix}
                    \mathrm{i} \\
                     - \mathrm{i}
                \end{pmatrix}\) 
            }
            \\
        
        &&\\
            Eigenvalues &
            \({k}_{-}^{2} = \left( \frac{\Omega}{2c} \right)^{2}\left( 1 + \chi_{0}^{*} \right);\ k_{+}^{2} = \left( \frac{\Omega}{2c} \right)^{2}\left( 1 + \chi_{0} \right)\)
            & \({k}_{\pm}^{2} = \left( \frac{\Omega}{2c} \right)^{2}\left( 1 + \chi_{0} \pm \frac{1}{2}\chi_{1} \right)\) \\
        
        &&\\
        \hline
    \end{tabular}
    \bigskip
    \caption{\textbf{Exemplar} \({\hat{\mathbf{K}}}_{\mathbf{\delta=0}}^{\mathbf{2}}\) \textbf{Matrices:} Two examples of \({\hat{K}}_{\delta=0}^{2}\) matrices, with eigenpulses with either broken or un--broken \(\mathcal{RC}\)-symmetry. i) Any static material with loss spontaneously breaks \(\mathcal{RC}\)-symmetry. It is the simplest example of a \(\mathcal{RC}\)-broken system, with eigenvectors peaked at a single frequency, illustrating the lack of inter-frequency coupling. ii) When loss is removed, \({\hat{K}}_{\delta=0}^{2}\) becomes a scalar function of the Pauli matrix \(\sigma_{x}\), and thus its eigenvectors are real.}
    \label{tab: ExemplarK2Matrices}
\end{table}

With example eigenvectors obtained, solutions to the wave equation can then be simply found. In the case of \(\mathcal{RC}\)-broken symmetry, Table \ref{tab: ExemplarK2Matrices} shows the waves take the textbook form of

\begin{equation}
    \psi_{B}\mathfrak{= R}\left\{ A_{+}e^{\mathrm{i}\frac{\Omega}{2}\left( \frac{z}{c}\sqrt{1 + \chi_{0}} - t \right)} \right\}\mathfrak{+ R}\left\{ A_{-}e^{\mathrm{i}\frac{\Omega}{2}\left( - \frac{z}{c}\sqrt{1 + \chi_{0}} - t \right)} \right\}.
    \label{eq: RCBrokenWaveSolution}
\end{equation}

This equation is well understood, with the key feature that the complex nature of the time-domain eigenpulses means that real waves are only achieved by coupling time and space into travelling waves of the form \(f\left( \pm \frac{z}{c}\sqrt{1 + \chi_{0}} - t \right)\). This can be contrasted with \(\mathcal{RC}\)-unbroken waves which from Table \ref{tab: ExemplarK2Matrices} are,

\begin{equation}
    \psi_{U} = \left| A_{+} \right|\cos\left( {\overline{k}}_{0, +}z + \phi_{+} \right)\cos\left( \frac{\Omega}{2}t \right) + \left| A_{-} \right|\cos\left( {\overline{k}}_{0, -}z + \phi_{-} \right)\sin\left( \frac{\Omega}{2}t \right).
    \label{eq: RCUnbrokenWaveSolution}
\end{equation}

The coupling between spatial and temporal parts of the wave, which makes it possible to split \(\psi_{B}\) into right- and left-moving parts, is no longer present for \(\psi_{U}\), and thus there is no defined direction of propagation.

\begin{figure}
    \centering
    \includegraphics[width=0.9\textwidth]{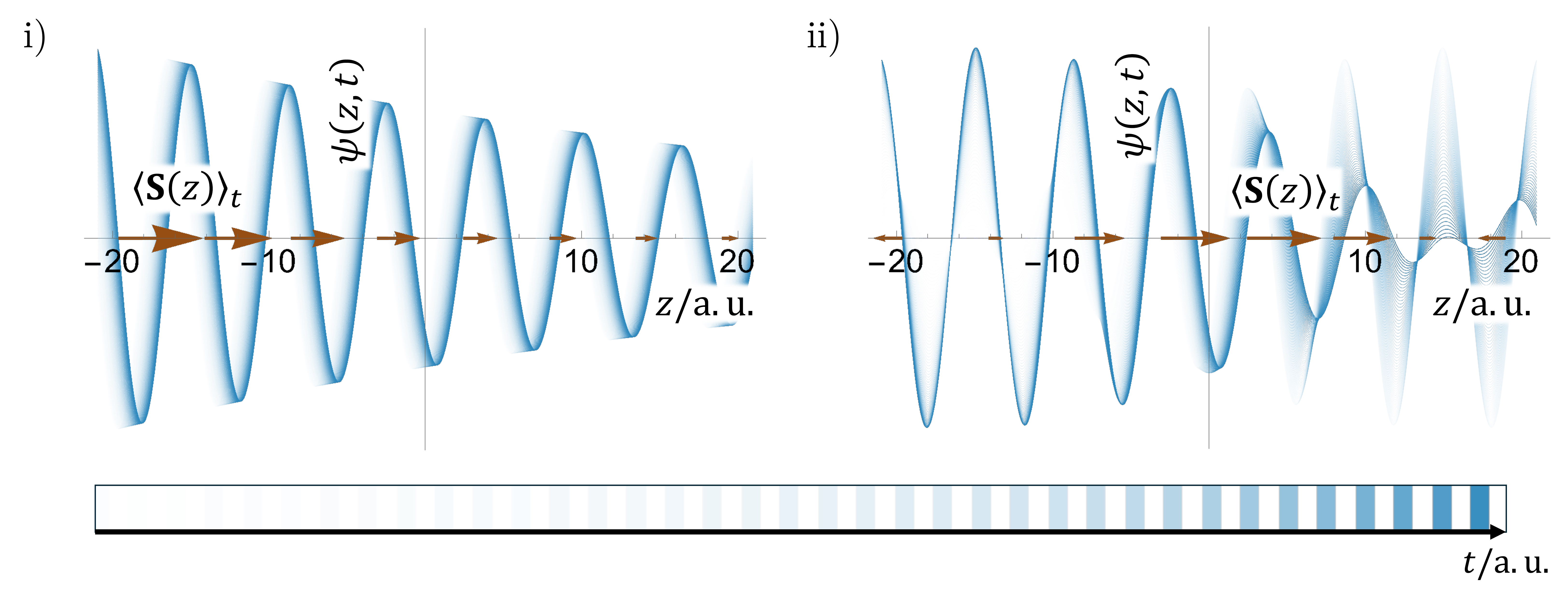}
    \caption{\textbf{Directionality of Waves:} Waves with different \(\mathcal{RC}\)-symmetry properties, as a function of \(z\), with their time dependence shown through exponential fading of the wave. The time-averaged Poynting vector is shown with arrows along the \(z\)-axis. Wavenumbers chosen for visualisation purposes only. i) A wave \(\left({k}_{\pm}=1\pm\frac{\mathrm{i}}{50}\right)\) propagating through a non-time-varying lossy medium, as an example of \(\mathcal{RC}\)-symmetry-broken waves. Notably, the wave points in a well-defined direction everywhere. ii) An \(\mathcal{RC}\)-unbroken wave \(\left({k}_{\pm}=1\pm\frac{1}{20}\right)\) travelling through a time-varying medium. The Poynting vector repeatedly swaps direction in space so that the wave cannot be meaningfully said to be travelling in any given direction.}
    \label{fig: WaveDirectionality}
\end{figure}

This can be elucidated in the limit of \(1 + \chi_{0} \ll \chi_{1}\), where the difference between the two wavenumbers is small. In this limit, both \(\left| A_{\pm} \right|\) waves appear to evolve with the same wavenumber, but with some spatially varying relative phase. Since it's possible to build a propagating wave from two spatially and temporally coherent beams by adjusting their relative phases, locally \(\psi_{U}\) may appear to propagate in a particular direction. However, over long enough distances, the spatial variation in the relative phase rewrites this local phase relationship. Thus, a locally right-moving wave at one point appears as a standing wave some distance away, and as a left-moving wave, even further away. This fact directly implies that it is fundamentally impossible to assign causal directions to \(\mathcal{RC}\)-unbroken waves (Fig. \ref{fig: WaveDirectionality}): directionality is something that can only be defined for \(\mathcal{RC}\)-broken waves.

As discussed earlier, intuition for these phenomena can be gleaned through analogies to spatial crystals. At the band-edge of a spatially periodic medium, repeated spatial reflections ensure that waves at the boundary are standing waves in space. These possess intensity maxima and minima spatially aligned with peaks and troughs of the perturbing susceptibility, and appropriately perturbed frequencies. The \(\mathcal{RC}\)-unbroken waves described above are perfect temporal analogues of this: they form standing waves in time; their wavenumbers are perturbed based on how their electric fields align with the applied susceptibility; and information about their initial direction is lost through repeated temporal reflections.

\section{Transmission Infinities in a Time-Varying Slab}\label{transmission-infinities-in-a-time-varying-slab}

Since \(\mathcal{RC}\)-symmetry is simply a feature of a system's response in time, the potential for \(\mathcal{RC}\)-unbroken phenomena exists more broadly than for the infinite medium considered above. In this section, we focus on a time-varying slab, where time-variations can overcome radiative losses from the ends of the slab. This can create a stable mode without any incident field, thus resulting in divergent transmission and reflection operators.

The operator formalism\cite{horsley2023eigenpulses} followed in this paper allows the ready calculation of the transmission operator. Following the standard non-time-varying derivation of the transmission coefficient, with a little additional care regarding operator ordering, we arrive at the following formula for the transmission operator,

\begin{equation}
    \hat{t} = \hat{n}\hat{\widetilde{F}}^{- 1},
    \label{eq: TransmissionOperatorDefinition}
\end{equation}
where the Fabry--Perot--like operator $\hat{\tilde{F}}$ is given by
\begin{equation}
    \hat{\widetilde{F}} = \left( \frac{(1 + \hat{n})}{2}e^{- \mathrm{i}{\hat{K}}_{\delta=0}L}\frac{(1 + \hat{n})}{2} - \frac{(1 - \hat{n})}{2}e^{\mathrm{i}{\hat{K}}_{\delta=0}L}\frac{(1 - \hat{n})}{2} \right),
    \label{eq: FabryPerotFactorOperatorDefinition}
\end{equation}
where \(L\) is the length of the cavity, and the refractive index operator is given by
\begin{equation}
    \hat{n} = \left( \frac{\omega}{c} \right)^{- 1}{\hat{K}}_{\delta=0}.
    \label{eq: RefractiveIndexOperatorDefintion}
\end{equation}

The lack of a defined direction in \(\mathcal{RC}\)-unbroken waves has a significant impact on the definition of \(\hat{n}\). In the case of static media, a square root of \(K^{2}\) can always be chosen to ensure that \(n\) is \(\mathcal{RC}\)-symmetric. This condition is equivalent to ensuring that \(n\) corresponds to a real-valued time-domain response. However, in \(\mathcal{RC}\)-unbroken media, \(K^{2}\) can no longer be chosen to ensure this. Thus, the standard semi-infinite transmission \(\left( \frac{n_{1}}{n_{1} + n_{2}} \right)\) and reflection \(\left( \frac{n_{1} - n_{2}}{n_{1} + n_{2}} \right)\) coefficients are \emph{also} no longer real time-domain operators. Physically, this arises from the fact that semi-infinite reflection coefficients require that a definite `outgoing' wave be well defined in both media, which as discussed above cannot be defined for media with \(\mathcal{RC}\)-unbroken waves!  This can equivalently be seen as a result of loss protection. Due to this, any continuous waves emitted infinitely deep in the time-varying medium are still guaranteed to reach the boundary with sensitivity to their initial conditions. By contrast, \(\hat{t}\) for the complete slab can be easily verified as \(\mathcal{RC}\)-symmetric, since it treats outgoing waves in the vacuum where they are well-defined.

For this discussion, we caution against interpreting this transmission coefficient in terms of a series of successive reflections from the (spatial) slab boundaries, as is the standard derivation for transmission through a static 1D slab. Whilst the same infinite sum can be performed, the fields corresponding to each successive reflection are not \(\mathcal{RC}\)-symmetric, and thus not real in the time-domain. This should perhaps not be surprising; any multi-reflection picture in the time-domain must account for temporal reflections, as well as reflections from the spatial boundaries.

\begin{figure}
    \centering
    \includegraphics[width=1\textwidth]{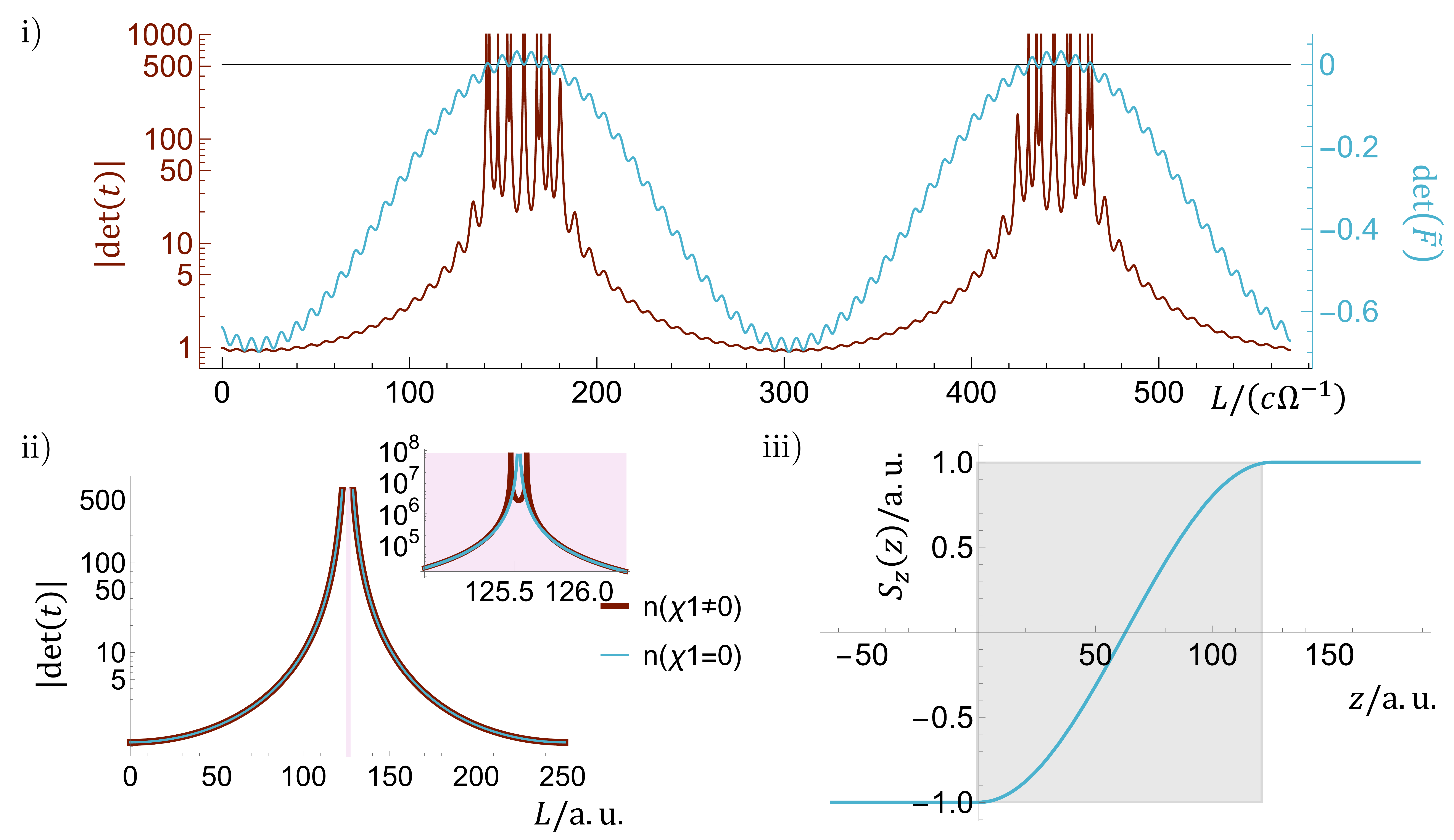}
    \caption{\textbf{Transmission Infinities:} i) Determinant of the transmission operator (brown) and reduced Fabry-Perot factor (cyan). From (\ref{eq: TransmissionOperatorDefinition}), infinities in \(\det(t)\) must arise from zeroes in \(\det\left( \widetilde{F} \right)\), which is here seen to repeatedly cross \(0\) whilst remaining real. ii) Magnitude of the transmission operator determinant for an example time-varying slab (\(\chi_{0}=0\), \(\chi_{1}=0.1\)) of length \(L\), with and without reflections at the boundary of a slab (the thick brown, and thin cyan, lines, respectively). Inset: The same functions, plotted close to their singularities. iii) Normalised local Poynting vector of the cavity mode at a transmission pole, here with vanishing boundary reflections.}
    \label{fig: TransmissionInfinities}
\end{figure}

The transmission operator (\ref{eq: TransmissionOperatorDefinition}) in systems supporting \(\mathcal{RC}\)-unbroken waves can still be evaluated as a function of the cavity length \(L\) (Fig. \ref{fig: TransmissionInfinities}.i). For small cavity lengths, this matches equivalent results for standard Fabry-Perot cavities: perfect transmission occurs periodically. However, as the cavity length increases, the transmission coefficient has eigenvalues which steadily increase above \(1\), eventually diverging. This divergence occurs at the longest reasonable length scale in the system, occurring at cavity lengths \(L\sim\frac{\pi}{{{k}}_{+} - {{k}}_{-}}\), a fact confirmed by explicit calculations of \(\hat{\widetilde{F}}\).

The physical origin of these poles can be made clear by reducing the permittivity contrast between the time-varying medium and its surroundings. Setting \(\chi_{0} = 0\) is achieved easily. However, as \(\chi_{1} \rightarrow 0\), the cavity length associated with the first divergent transmission coefficient also diverges. Thus, for simplicity, we consider setting \(\chi_{1} \rightarrow 0\) in \(n\), whilst \({\hat{K}}_{\delta=0}L\) is held constant (Fig. \ref{fig: TransmissionInfinities}.ii). In this limit,

\begin{equation}
    \hat{n} = \begin{pmatrix}
        - 1 & 0 \\
        0 & 1
    \end{pmatrix}.
    \label{eq: LimitingRefractiveIndex}
\end{equation}

The implication of this on the reduced Fabry-Perot factor is significant. The factors of \(\frac{(1 \pm \hat{n})}{2}\) reduce to projection operators \(\hat{\Pi}_{\pm}\) onto frequencies with \(\pm \omega > 0\), reducing \(\hat{\widetilde{F}}\) to a diagonal matrix. Thus, \(\det\left( \hat{\widetilde{F}} \right) = 0\) whenever both of the following conditions hold,

\begin{equation}
    \hat{\Pi}_{\pm}e^{\mp \mathrm{i}{\hat{K}}L}\hat{\Pi}_{\pm} = 0.
    \label{eq: CavityModeCondition}
\end{equation}

Physically, this means that transmission diverges precisely when propagation through the cavity acts to map a state of positive frequency to one with negative frequency, or vice versa (in fact, in this limit, both occur simultaneously). Thus, a wave which appears to move rightwards at one boundary is mapped to a leftwards-moving wave at the other boundary, forming a stable mode which continually emits energy to its surroundings. This matches precisely calculations of the microscopic Poynting vector in such a slab (Fig. \ref{fig: TransmissionInfinities}.iii). These modes can be constructed with any phase. However, when \(\chi_{1}\) is again increased, reflections from the boundary become phase sensitive, splitting the cavity modes, and thus the pole in the transmission operator.

\section{Conclusion}\label{conclusion}

Using the operator formalism for describing wave propagation in dispersive, time varying materials we have taken a truncated range of frequencies, finding a range of symmetry-protected phenomena that should be present in a wide range of systems. These phenomena include the existence of dissipation--free waves in lossy time--modulated systems that arise from \(\mathcal{RC}\)--symmetry (combined conjugation and frequency inversion) -- waves which also have no defined direction of propagation.  Connected with this, we showed the existence of divergent transmission in a slab of lossy time--modulated material, provided the length of the slab is tuned to a particular value.  Interestingly, this divergence remains even when there are no boundary reflections confining the field.

A more practical result is that, just as there is no minimum potential required to create a bandgap in electron dispersion, there's no minimum modulation strength required to achieve effects like maintaining a cavity mode amplitude with temporal reflections; if a system is lossless, arbitrarily low susceptibility variations can be used, at the expense of a linear increase in system length. In fact, the nonlinear crystals used in optical parametric oscillators present a perfect material for such tests, for a pump beam strong enough to be essentially unaffected by the probe. Conversely, even in systems with high loss, sufficient modulation can still lead to the lossless modes we identify here -- including evanescent waves in strongly modulated metallic systems.

\section*{Acknowledgements}

CMH thanks Oliver Breach, James Walkling, and Gregory Chaplain for useful discussions, and acknowledges financial support from the Engineering and Physical Sciences Research Council (EPSRC) of the UK via the Exeter University Physics DTP. IRH and SARH acknowledge financial support from the EPSRC via the META4D Programme Grant (EP/Y015673/1).  SARH thanks the Royal Society and TATA for financial support (RPG-2016-186). JRC acknowledges support from DSTL.

\bibliography{apssamp}

\appendix

\section{Choosing an Operator Square Root}\label{square-root-choice}

To make explicitly clear that all of the solutions to the wave equation are encoded by \ref{eq: OperatorFormWaveSolution}, we consider solutions to the wave equation written in terms of the eigenvectors and eigenvalues of \(K^{2}\), defined as \(K^{2}\mathbf{e}_{i} = k_{i}^{2}\mathbf{e}_{i}\). The solutions to the wave equation are thus found simply as

\begin{equation}
    \widetilde{\psi} = \sum_{i}^{}{\frac{1}{2}\left( A_{i, +}e^{\mathrm{i}k_{i}z} + A_{i, -}e^{- \mathrm{i}k_{i}z} \right)\mathbf{e}_{i}}.
    \label{eq: BasicEigendecompositionWaveSolution}
\end{equation}

This formulation makes an important point clear. When evaluating \(\sqrt{K^{2}}\), there are an infinite set of possible roots which could be chosen, each distinct in the choice of sign given to each eigenvalue. However, writing \(\widetilde{\psi}\) in this form demonstrates that such a choice is inconsequential: swapping the sign of the eigenvalue \(k_{i}\) simply swaps the roles of \(A_{\pm}\).

\section{Operator Definitions}\label{operator-definitions}

Noting that coupling only occurs within a given frequency ladder, it follows that \(K^{2}\) is block diagonal, with blocks \(K_{\delta}^{2}\). Explicitly evaluating \ref{eq: FrequencyDomainWaveEquation} within each of these blocks allows \(K_{\delta}^{2}\) to be written in terms of the infinite-dimensional, but now at least discrete, matrices

\begin{equation}
    \omega_{\delta} = \Omega
    \begin{pmatrix}
        \ddots & \ddots & \vdots & \vdots & \vdots & \iddots \\
        \ddots & - \frac{3}{2} + \delta & 0 & 0 & 0 & \cdots \\
        \cdots & 0 & - \frac{1}{2} + \delta & 0 & 0 & \cdots \\
        \cdots & 0 & 0 & \frac{1}{2} + \delta & 0 & \cdots \\
        \cdots & 0 & 0 & 0 & \frac{3}{2} + \delta & \ddots \\
        \iddots & \vdots & \vdots & \vdots & \ddots & \ddots 
    \end{pmatrix},
    \label{eq: BlockFrequencyOperator}
\end{equation}

\begin{equation}
    \hat{C}_{\delta} = \frac{1}{2}
    \begin{pmatrix}
        \ddots & \ddots & \ddots & \vdots & \vdots & \iddots \\
        \ddots & 0 & 1 & 0 & 0 & \cdots \\
        \ddots & 1 & 0 & 1 & 0 & \cdots \\
        \cdots & 0 & 1 & 0 & 1 & \ddots \\
        \cdots & 0 & 0 & 1 & 0 & \ddots \\
        \iddots & \vdots & \vdots & \ddots & \ddots & \ddots 
    \end{pmatrix},
    \label{eq: BlockCouplingOperator}
\end{equation}

such that

\begin{equation}
    \hat{K}_{\delta}^{2} = \frac{\omega_{\delta}^{2}}{c^{2}}\left( 1 + \chi_{0}\left( \omega_{\delta} \right) + \hat{C}_{\delta}\chi_{1}\left( \omega_{\delta} \right) \right).
\end{equation}

\section{RC-Symmetry}\label{rc-symmetry}

In most physical systems, waves are real-valued. In the time-domain, this constrains that the wave equation contains only real functions of derivatives. In the frequency domain, this becomes the statement that \(\chi( - \omega) = \chi^{*}(\omega)\). In this section, we generalise this symmetry to the matrices describing time-varying materials, and show what this symmetry implies about the eigenvectors and eigenvalues of \(K^{2}\).

We define operators for frequency reflection (\(\mathcal{R}\)) and complex conjugation (\(\mathcal{C}\)) as \(\mathcal{R}\widetilde{\psi}(\omega) = \widetilde{\psi}( - \omega)\), and \(\mathcal{C}\widetilde{\psi}(\omega) = \left( \widetilde{\psi}(\omega) \right)^{*}\mathcal{C}\). The Fourier Transform of a real function possesses the symmetry, henceforth referred to as \(\mathcal{RC}\)-symmetry of \(\mathcal{RC}\widetilde{\psi}\mathcal{C =}\widetilde{\psi}\).

Applying this operator to the frequency domain wave equation (\ref{eq: FrequencyDomainWaveEquation}) gives

\begin{equation}
    \frac{\partial^{2}\mathcal{RC}\widetilde{\psi}\mathcal{C}}{\partial z^{2}}\mathcal{= - RC}K^{2}\widetilde{\psi}\mathcal{C.}
    \label{eq: RCWaveEquation}
\end{equation}

Noting that \(\frac{\partial^{2}\mathcal{RC}\widetilde{\psi}\mathcal{C}}{\partial z^{2}} = \frac{\partial^{2}\widetilde{\psi}}{\partial z^{2}} = - K^{2}\widetilde{\psi}\), and using the self-inverse nature of \(\mathcal{C}\), one obtains

\begin{equation}
    K^{2}\widetilde{\psi}\mathcal{= RC}K^{2}\mathcal{RC}\widetilde{\psi}.
    \label{eq: RCK2Condition}
\end{equation}

Which must hold for all \(\mathcal{RC}\)-symmetric \(\widetilde{\psi}\). However, since \(K^{2}\) commutes with complex scalars (unlike \(\mathcal{C}\)), this in fact holds in general, and thus \(\mathcal{RC}K^{2}\mathcal{RC} = K^{2}\).

Following similar calculations in the study of \(\mathcal{PT}\)-symmetry\cite{bender2007making}, we can investigate the implications of this symmetry on the eigenvectors and eigenvalues of \(K^{2}\). In the argument below, we assume \(K^{2}\) contains a spectrum of distinct eigenvalues, but it is generalised without significant difficulty. Let the eigenvectors and eigenvalues be defined by \(K^{2}\mathbf{e}_{i} = k_{i}^{2}\mathbf{e}_{i}\). Then,

\begin{equation}
    \mathcal{RC}K^{2}\mathcal{RC}\mathbf{e}_{i} = k_{i}^{2}\mathbf{e}_{i}.
    \label{eq: RCEigendecompositionEquation}
\end{equation}

Operating on both sides with \(\mathcal{RC \cdot C}\), one obtains an eigenvalue equation:

\begin{equation}
    K^{2}\left( \mathcal{RC}\mathbf{e}_{i}\mathcal{C} \right) = k_{i}^{*2}\left( \mathcal{RC}\mathbf{e}_{i}\mathcal{C} \right).
    \label{eq: NewRCEigenvectorEquation}
\end{equation}

It thus follows that, \(\mathcal{RC}\mathbf{e}_{i}\mathcal{C}\) is an eigenvector with eigenvalue \(k_{i}^{*2}\). The implications of this differ depending on whether \(k_{i}^{2}\) is real. When it is, by the assumption of distinct eigenvalues, this implies \(\mathcal{RC}\mathbf{e}_{i}\mathcal{C} \propto \mathbf{e}_{i}\). Otherwise, \(\mathcal{RC}\mathbf{e}_{i}\mathcal{C}\) must be a new eigenvector. Thus, the pair, \(\left\{ \left( \mathbf{e}_{i},k_{i}^{2} \right),\left( \mathcal{RC}\mathbf{e}_{i}\mathcal{C,}k_{i}^{*2} \right) \right\}\) obeys \(\mathcal{RC}\)-symmetry, whilst the eigenvectors fail to do so alone.

This information is summarised in Table~\ref{tab: RCSymmetryTable}. Statements in each column are equivalent, so \(\mathcal{RC}\)-symmetry-breaking can be identified by eigenvalue alone; frequently more convenient than explicitly finding the eigenvectors of \(K^{2}\).

\begin{table}[H]
    \centering
    \begin{tabular}{|c|c|c|}
        \hline
             &
            \(\mathcal{RC}\)-unbroken. &
            \(\mathcal{RC}\)-broken. \\
        \hline
        
        && \\
            \begin{minipage}{0.15\textwidth}
                Eigenvectors \(\mathbf{e}_{i}\)
            \end{minipage} &
            \begin{minipage}{0.35\textwidth}
                \(\mathbf{e}_{i}\) can be shifted by a phase to ensure that \(\mathcal{RC}\mathbf{e}_{i}\mathcal{C =}\mathbf{e}_{i}\). This symmetry ensures that both positive and negative frequencies are of equal magnitude in \(\mathbf{e}_{i}\).
            \end{minipage} &
            \begin{minipage}{0.35\textwidth}
                No non-trivial scaling of \(\mathbf{e}_{i}\) is \(\mathcal{RC}\)-symmetric. Attempting to use the symmetry \(\mathcal{RC}\mathbf{e}_{i}\mathcal{C}\) returns a second eigenvector \({\overline{\mathbf{e}}}_{i}\) with a conjugated eigenvalue.
            \end{minipage} \\
            
        && \\
            \begin{minipage}{0.15\textwidth}
                Eigenvalues \(k_{i}^{2}\)
            \end{minipage} &
            \begin{minipage}{0.35\textwidth}
                \(k_{i}^{2}\) is real.
            \end{minipage} &
            \begin{minipage}{0.35\textwidth}
                \(k_{i}^{2}\) is complex. Conjugating it returns another eigenvalue of \(K^{2}\).
            \end{minipage} \\
        
        && \\
        \hline
    \end{tabular}
    \bigskip
    \caption{\textbf{Classifying} \(\hat{\mathbf{K}}^{\mathbf{2}}\) \textbf{Eigenpulses:} The eigenpulses of any \(\mathcal{RC}\)-symmetric matrix without repeated eigenvalues can be identified according to the table above. These symmetries are generalised to degenerate \(\hat{K}^{2}\) by replacing eigenvectors with their corresponding degenerate subspaces.}
    \label{tab: RCSymmetryTable}
\end{table}

With this in mind, the eigenvector/eigenvalue indices in (\ref{eq: BasicEigendecompositionWaveSolution}) can be split into symmetry-broken/unbroken sets (\(n_{B}\)/\(n_{U}\)), allowing \(\widetilde{\psi}\) to be made explicitly \(\mathcal{RC}\)-symmetric;

\[\widetilde{\psi} = \sum_{i \in n_{U}}^{}{\frac{1}{2}\left( A_{i}e^{ik_{i}z} + A_{i}^{*}e^{- ik_{i}z} \right)\mathbf{e}_{i}} + \sum_{i \in n_{B}}^{}{\frac{1}{2}\left( A_{i}e^{ik_{i}z}\mathbf{e}_{i} + A_{i}^{*}e^{- ik_{i}^{*}z}{\overline{\mathbf{e}}}_{i} \right)}.\]

Returning to the time-domain, this becomes explicitly real.

\[\psi = \sum_{i \in n_{U}}^{}{\left| A_{i} \right|\cos\left( k_{i}z + \arg\left( A_{i} \right) \right)e_{i}(t)} + \sum_{i \in n_{B}}^{}{\left| A_{i} \right|\mathfrak{R}\left\{ e^{i\left( k_{i}z + \arg\left( A_{i} \right) \right)}e_{i}(t) \right\}},\]

where \(e_{i}(t)\) is obtained from the inverse Fourier Transform of \(\mathbf{e}_{i}\). \(e_{i}(t)\) is real for \(i \in n_{U}\), and complex for \(i \in n_{B}\). As in Table~\ref{tab: RCSymmetryTable}, the eigenvector \({\overline{\mathbf{e}}}_{i}\) is defined as \({\overline{\mathbf{e}}}_{i}\mathcal{\equiv RC}\mathbf{e}_{i}\mathcal{C}\).

The above decomposition hints at a qualitative difference between waves based on their symmetry. However, whilst it is easy to find examples of \(\mathcal{RC}\)-broken eigenpulses (any eigenpulses in non-time-varying lossy materials, for instance), it is not immediately obvious that \(\mathcal{RC}\)-\emph{un}broken eigenpulses should ever be found physically, aside from a single trivial eigenpulse for \(|\delta| = 0.5\) (in optics, this corresponds to a static \(\mathbf{D}\)-field). However, the \(\frac{\omega_{|\delta| = 0.5}^{2}}{c^{2}}\) prefactor on \(\hat{K}_{|\delta| = 0.5}^{2}\) prevents further eigenpulses coupling between positive and negative frequencies, thus preventing the existence of any further \(\mathcal{RC}\)-unbroken eigenpulses associated within \(|\delta| = 0.5\). Since \(\mathcal{RC}\)-symmetric eigenpulses can only exist on \(\mathcal{RC}\)-symmetric ladders, their only possible remaining origin is \(\hat{K}_{\delta=0}^{2}\).

\end{document}